\documentclass[sigplan]{acmart}
\renewcommand\footnotetextcopyrightpermission[1]{} 

\makeatletter
\DeclareRobustCommand\ttfamily
        {\not@math@alphabet\ttfamily\mathtt
         \fontfamily\ttdefault\footnotesize\selectfont}
\makeatother

\startPage{1}

\setcopyright{none}

\usepackage{booktabs}   
\usepackage{subcaption} 
\usepackage{multirow}
\usepackage{makecell}
\usepackage{listings}
\usepackage{amsthm}
\theoremstyle{definition}

\definecolor{mpcolor}{rgb}{0.1,0.9,0.1}
\definecolor{fwcolor}{rgb}{0.0,0.0,1.0}
\definecolor{rmcolor}{rgb}{0.7,0.0,0.0}

\newcommand{\Name}{LoopLearner}

\begin{document}

\title[LoopLearner]{Learning to Make Compiler Optimizations \\ More Effective}


\author{Rahim Mammadli}
\affiliation{
  \department{Department of Computer Science}
  \institution{Technical University of Darmstadt}
}
\email{rahim.mammadli@tu-darmstadt.de}

\author{Marija Selakovic}
\affiliation{
  \department{Department of Computer Science}
  \institution{Technical University of Darmstadt}
}
\email{m.selakovic89@gmail.com}

\author{Felix Wolf}
\affiliation{
  \department{Department of Computer Science}
  \institution{Technical University of Darmstadt}
}
\email{felix.wolf@tu-darmstadt.de}

\author{Michael Pradel}
\affiliation{
  \department{Department of Computer Science}
  \institution{University of Stuttgart}
}
\email{michael@binaervarianz.de}

\begin{abstract}
  Because loops execute their body many times, compiler developers
  place much emphasis on their optimization. Nevertheless, in view of
  highly diverse source code and hardware, compilers still struggle to
  produce optimal target code. The sheer number of possible loop
  optimizations, including their combinations, exacerbates the problem
  further. Today's compilers use hard-coded heuristics to decide when,
  whether, and which of a limited set of optimizations to apply. Often, this leads to highly unstable behavior, making the
  success of compiler optimizations dependent on the precise way a
  loop has been written. This paper presents \Name{}, which addresses the problem of compiler instability by predicting which way of writing a loop will lead to
  efficient compiled code. To this end, we train a neural network to
  find semantically invariant source-level transformations for loops
  that help the compiler generate more efficient code. Our model
  learns to extract useful features from the raw source code and
  predicts the speedup that a given transformation is likely to
  yield.
  We evaluate \Name{} with 1,895 loops from various performance-relevant benchmarks.
  Applying the transformations that our model deems most
  favorable prior to compilation yields an average speedup of
  1.14x. When trying the top-3 suggested transformations, the average speedup even
  increases to 1.29x.
  Comparing the approach with an exhaustive search through all available code transformations shows that \Name{} helps to identify the most beneficial transformations in several orders of magnitude less time.
\end{abstract}

\ccsdesc[500]{Software and its engineering~Compilers}
\ccsdesc[500]{Computing methodologies~Neural networks}
\ccsdesc[300]{Computing methodologies~Supervised learning by regression}


\maketitle
\pagestyle{plain} 

\section{Introduction}


The optimization techniques used in modern compilers are continuously improving. In view of the increasing complexity of hardware and software, the effectiveness of compiler optimizations
becomes crucial in achieving satisfactory system performance. However, despite the tremendous progress of compiler technology, the optimizations a compiler applies are usually limited to a fixed set of program transformations. Furthermore, compiler developers manually design optimization heuristics that control program compilation and optimization. Writing these heuristics requires expert knowledge and is one of the most difficult and time-consuming tasks in compiler development. This is why compiler optimizations are not guaranteed to produce optimal output, and in fact, they may even degrade performance in some cases.

A recent study by \citet{Gong:2018} illustrates the challenges compiler developers face today. Looking at how source-level loop transformations affect performance, the authors observed
that compilers are not only far from producing optimal code, but are also highly unstable: given semantically equivalent variants of the same piece of code, compilers produce target code that differs significantly in terms of performance. As a result of this ``compiler instability'', as Gong et al. named the problem, programmers are left without any guidance as to which variant of the source code to feed into the compiler.
To maximize performance, a programmer may choose to deal with compiler instability by (a) systematically trying as many semantically equivalent code variants as possible and measure which performs best, or (b) learning through experience which variant works best for a given compiler. Since the first option is very time consuming and the second option requires expert knowledge of the underlying compiler, both strategies are of limited use in practice.

To mitigate the problem of compiler instability, we present \emph{\Name{}}, a learning-based approach that predicts semantics-preserving transformations of a loop that will improve the performance of the compiled program. Given a loop and a search space of such transformations, \Name{} predicts which transformation or sequence of transformations will yield the best-performing target code with a given compiler.
The search space explored by \Name{} consists of around 3,000 sequences of transformations, composed of five basic optimizations, their combinations, and different parametrizations.
We focus on loops for two reasons. First, optimizing loops is important because the loop body is repeatedly executed, not seldom thousands of times, which in total accounts for a significant fraction of the overall execution time. Second, loop transformations are one of the major optimizations supported by modern compilers, which is why loops are at the core of compiler instability.

We envision \Name{} to be useful in multiple scenarios. First, it can assist developers in deciding how to write a loop. By predicting which variant of a loop yields the best performance, developers can make an informed decision, instead of relying on their intuition. Second, the approach can guide an automated pre-processing step that applies code transformations before handing the code over to the compiler. Such pre-processing does not require any developer attention and mitigates the problem of compiler instability without the need to change the compiler itself.
And, of course, one could also integrate our predictive model directly into the compiler to improve its stability.
In the second and third usage scenario, \Name{}'s predictions complement the built-in optimization heuristics of the compiler by presenting the code in a way that will make best use of these heuristics.

We define the problem of predicting the best transformation for a loop as a regression problem: based on the source code of a given loop, \Name{} learns to predict the speedup that a certain transformation is likely to yield. After training the model with tens of thousands of examples, we query the model for each transformation to determine which one gives the highest performance improvement.
To effectively learn the performance benefits of transformations on specific code, we need a suitable encoding of both inputs.
\Name{} encodes source code as a sequence of tokens, and we compare different representations of individual tokens.
To encode transformations, we present a novel, compact representation that ensures that similar transformations have a similar representation.
\Name{} uses a convolutional neural network architecture, which has been proven as very effective on compositional data.

One of the key challenges in choosing among the available code optimizations is the large space of possible transformations.
A naive approach could apply each transformation, then run the compiled code, and measure its execution time.
Unfortunately, this approach takes significant time, in particular, because reliable performance measurements require executing the code repeatedly.
Instead of executing transformed code, \Name{} queries a predictive model once per transformation.
Since querying our neural model is very fast and because queries for different transformations can be run in batches, our approach reduces the effort for finding a suitable transformation by multiple orders of magnitude.


Prior learning-based work on improving optimizing compilers aims at finding suitable compiler heuristics, including the work by \citet{Yuki:2010}, who predict optimal loop tiling sizes, \citet{Stephenson:2005}, who determine the best loop unrolling factor, and \citet{Simon:2013}, who construct compiler heuristics automatically.
Our approach differs those approaches in several ways.
One difference is that we consider a much larger space of optimizations, that is, nearly 3,000 combinations of five common loop optimizations---unrolling, unroll and jam, tiling, distribution, and interchange, including variations of their parameters. Another distinctive feature of our approach is that it reasons about source-level transformations to be applied before passing a program to the compiler, instead of optimization decisions taken in the compiler. Finally, \Name{} involves neither the manual design nor the pre-selection of any features. Instead, we feed the source code as-is into a neural network that learns how to identify suitable features on its own.
\citet{Cummins2017a} also train a neural model that predicts from raw code how to support code optimization. However, their model focuses on a small set of optimization parameters used in the compiler, e.g., whether to map a kernel to the CPU or the GPU, whereas we consider a larger space of transformations applied before passing code to the compiler.

To evaluate \Name{} we use an extensive collection of nested loops from the empirical study by \citet{Gong:2018}. To train the model, we consider all transformations the study used to create loop mutations. In total, the data set amounts to around 70,000 data points, originating from 1,895 unique loops from 18 benchmarks and almost 3,000 unique transformations. One transformation consists of a sequence of one or more loop transformations and their parameters. We find that our model has a precision of 73\% when predicting speedups. Furthermore, by ranking all transformations based on their predicted performance improvements and by applying the top-1 transformation, \Name{} achieves a speedup of 1.14x, on average across all loops. If the developer or tool tries the top-3 suggested transformations and picks the best one, the average speedup increases even to 1.29x.

In summary, this paper makes the following contributions:
\begin{itemize}
	\item \emph{Learning-based approach to mitigate compiler instability}.
	We are the first to systematically mitigate the problem of compiler instability through a learned model that predicts source-to-source transformations likely to make compiler optimizations more effective.
	The deep learning-based model automatically extracts features from a given loop, without any manual feature engineering.

	\item \emph{Search space}. The approach scales to a large search space consisting of thousands of transformations. The search space is built from five common and semantically invariant loop transformations, applied alone or in sequence, and their several parameters.

	\item \emph{Empirical evidence}. We empirically demonstrate that applying the transformation our model deems most favorable yields an average speedup of 1.14x (for the best predicted transformation) or 1.29x (when considering the top-3 predictions).

\end{itemize}

The remainder of this paper is organized as follows. Section~\ref{sec:background} summarizes the problem of compiler instability described by \citet{Gong:2018}. Section~\ref{sec:approach} presents our approach to the selection of beneficial loop transformations. Section~\ref{sec:eval} discusses experimental settings and results. Finally, we discuss related work in Section~\ref{sec:related} and review our results in Section~\ref{sec:con}.

\section{Background}
\label{sec:background}

The attribute \emph{stable} characterizes a compiler that produces the
same performance for any semantically equivalent variant of a
program. In their study, \citet{Gong:2018} evaluate the
stability of modern compilers by applying several source-to-source
transformations to obtain semantically equivalent code variants and by
measuring the variation in their execution time. To illustrate the
effect of program transformations on compiler stability, consider the
example in Listing~\ref{lst:unrolling}. The first loop is extracted
from function \emph{Regclass} in the SPEC CPU2000 benchmark
suite. After unrolling the loop with a factor of two, yielding the
second loop in the listing, the Clang compiler generates output that is,
on average, 1.19x faster than the original loop.

\lstdefinestyle{my_style}{
  language=c,
  captionpos=b,
  xleftmargin=0.0cm,
  basicstyle=\ttfamily\scriptsize
}

\begin{center}
\begin{minipage}{\linewidth}
\begin{lstlisting}[caption={Original and unrolled loop in function \emph{Regclass} from the \emph{253.perlbmk} program in the SPEC CPU2000 benchmark suite.}, label={lst:unrolling}, style=my_style]
/* original loop */
for(Class = 0; Class < 256; ++Class){
    if(opnd[1 +(Class >> 3 & 31)] & 1 <<(Class & 7)){
         I32 cf = Perl_fold[Class];
         opnd[1 +(cf >> 3 & 31)] |= 1 <<(cf & 7);
    }
}

/* unrolled, factor = 2 */
for(Class = 0; Class <= 255; Class += 2) {
   if(opnd[1 +(Class >> 3 & 31)] & 1 <<(Class & 7)){
         I32 cf = Perl_fold[Class];
         opnd[1 +(cf >> 3 & 31)] |= 1 <<(cf & 7);
   }
   if(opnd[1 +(Class+1 >> 3 & 31)] & 1 <<(Class+1 & 7)){
         I32 cf = Perl_fold[Class+1];
         opnd[1 +(cf >> 3 & 31)] |= 1 <<(cf & 7);
   }
}

\end{lstlisting}
\end{minipage}
\end{center}

Gong et al. quantify compiler stability using the following two
metrics: \emph{intra-compiler} and \emph{inter-compiler
  stability}. The first metric, which is the focus of this paper, measures the stability of a single
compiler, while the second metric measures the stability across
multiple compilers. Although the authors of the study concede that
building a perfectly stable compiler is almost impossible, they show
that modern compilers have ample potential for improvement in this
direction. Specifically, they demonstrate that applying source-level
transformations prior to compilation can significantly reduce the
performance gap between variants of a loop.
A problem not addressed by prior work is which out of many possible transformations to apply to a given piece of code.

\begin{table}[t]
\centering
\caption{Loop transformations and their parameters.}
\begin{tabular}{ll}
\midrule
 Transformation & Parameters \\
 \midrule
 Unrolling & Unroll factor $\in \{2,4,8\}$\\
 Unroll-and-jam & Loop level, unroll factor $\in \{2,4\}$\\
 Tiling & Loop level, tile size $\in \{8,6,32\}$ \\
 Interchange & Lexicographical permutation number \\
 Distribution & No parameters \\
 \midrule

\end{tabular}
\centering

\label{tb:parameters}
\end{table}

The purpose of our work is to address the problem of
\emph{intra-compiler instability}, by learning code transformations
that should be applied to maximize the performance of the compiler
output. We train our model on the same source code examples and
transformations used in the original study by Gong et al. Each loop
transformation consists of a sequence of well-known base
transformations, which are listed in Table~\ref{tb:parameters}. To
ensure that transformations produce semantically equivalent output for
every loop, the space of considered transformations is limited to  sub-sequences of the following sequences:
\begin{itemize}
\item \emph{interchange} $\rightarrow$ \emph{unroll-and-jam} $\rightarrow$ \emph{distribution} $\rightarrow$ \emph{unrolling}
\item \emph{interchange} $\rightarrow$ \emph{tiling} $\rightarrow$ \emph{distribution} $\rightarrow$ \emph{unrolling}
\end{itemize}

In total, this space consists of almost 3,000 unique transformations (i.e.,
sub-sequences), each of them combining base transformations with
different parameters.
The number of transformations applied to a specific loop is much smaller (37, on average), because only some transformations can be applied in a semantics-preserving way.
Yet, as we show in Section~\ref{sub:comparison_with_exhaustive_search}, exhaustively exploring the performance impact of all transformations is still rather expensive.


\section{Approach}
\label{sec:approach}

In this section, we describe the \Name{} approach, which mitigates the problem of compiler instability by predicting loop transformations that enable the compiler to produce efficient target code.
We start with a rough overview and potential usage scenarios, before we define our learning problem. Afterwards, we discuss preprocessing steps applied to the data, before showing which encoding methods we experimented with. Next, we introduce our deep-neural-network (DNN) architecture and discuss design decisions made while building it. Finally, we specify the set of hyperparameters used to train the neural model.

\subsection{Overview}
\begin{figure*}[t]
\begin{center}
   \includegraphics[trim={0 0 0 1},clip,width=0.75\linewidth]{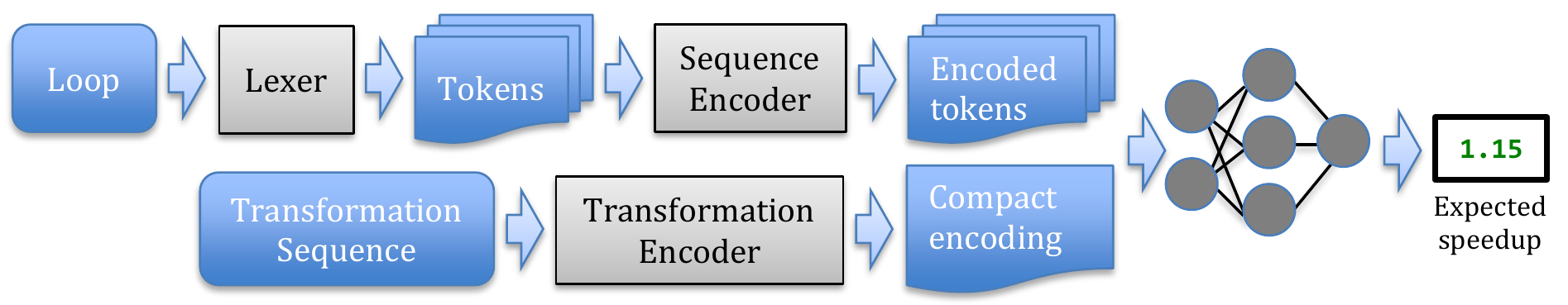}
\end{center}
   \caption{High-level overview of LoopLearner.}
\label{fig:approach_overview}
\end{figure*}

Figure~\ref{fig:approach_overview} illustrates our approach on a high level. The input to our network is a loop and a transformation that may be applied to it. We assume that the transformation is valid and does not affect the semantics of the program. For the dataset used in the evaluation, which we borrowed from Gong et al.~\cite{Gong:2018}, these properties are ensured using the polyhedral optimizer~\emph{Polyopt/C}~\footnote{\url{http://web.cse.ohio-state.edu/~pouchet.2/software/polyopt}} and the dependence analyzer~\emph{Candl}~\footnote{\url{http://icps.u-strasbg.fr/people/bastoul/public_html/development/candl}}.
As a first step, we tokenize the loop with the help of a lexer. The resulting sequence of tokens is then encoded using one of the methods discussed in Section~\ref{sub:codeEncoding}.
To feed the transformation into the model, the approach encodes it into a compact, similarity-preserving representation presented in Section~\ref{sub:transformationEncoding}.
Given both the code and the transformation, the model predicts the speedup, i.e., the ratio of the original loop's execution time divided by the execution time obtained by applying the transformation. Hence, having a set of valid transformations that can be applied to a given loop, our neural network can be used to rank them by their predicted speedup.
Given a ranked list of transformations, the user or a tool can then apply the transformation that is expected to produce the highest speedup.

\subsection{Interpreting Predictions}
To interpret the predictions of our model, we start by specifying a \textit{speedup threshold}, which is a hyperparameter used to classify the prediction as either advantageous, disadvantageous, or neutral. Formally, let $p$ be the prediction of the model, $a$ be the actual performance, and $t$ be a speedup threshold with $t>1$. Then, the prediction is assigned to one of three classes:
\begin{itemize}
  \item advantageous, if $p > t$
  \item disadvantageous, if $p < 1 - (t - 1)$
  \item neutral, if $ 1 - (t - 1) \le p \le t $
\end{itemize}
A prediction is considered to be accurate if:
$$(p > 1 \wedge a > 1)\vee (p \le 1 \wedge a \le 1) $$

Since our solution is intended to achieve speedup and avoid slowdown, we value a high precision rate for speedup predictions. Therefore, increasing $t$ (i.e., the range where the model predicts neutral) allows us to focus on clearer predictions of speedups and slowdowns, which is likely to increase precision but to reduce recall.

\subsection{Usage Scenarios}
\label{sub:usageScenarios}

A programmer or a tool facing the problem of choosing the best transformation for a given loop has multiple options. The first option involves applying no transformations and relying on the compiler to determine and apply the best set of optimizations. The second option is to test the performance of the loop with $k$ different transformations and choose the one producing the highest speedup. As discussed earlier, the number of transformations, their combinations, and the number of parameters that each of them accepts can result in a very high number of distinct transformations applicable to a given loop. Therefore, in most real-life scenarios measuring the performance of a loop with all possible transformations is not feasible. It can, however, be feasible to evaluate $k$ transformations if $k$ is a relatively small number.

To aid the programmer or tool in choosing the best set of transformations for a given loop we consider two usage scenarios of \Name{}:
\begin{itemize}
\item If evaluating the performance of loops with and without applying transformations is prohibitively expensive, we propose using LoopLearner in a \textit{static scenario}. This scenario implies applying the best advantageous transformation if such a transformation exists.
\item If evaluating the performance of up to $k$ mutations of loops is feasible, LoopLearner can be used in a \textit{dynamic scenario}, which involves applying the top-$k$ advantageous transformations and measuring their actual performance. If none of the transformations results in actual speedup, the original loop is left untouched. Otherwise, the programmer or a tool chooses the transformation resulting in the highest speedup.
\end{itemize}

\subsection{Definition of the Learning Problem}

The task of predicting a speedup achievable by applying a given transformation to the loop can be viewed as a regression problem. Specifically, given a dataset $\{(L_i, T_i) \to S_i\}_{i=1}^{N}$, where $N$ is the size of the dataset, $S_i$ is the speedup or slowdown resulting from applying the transformation $T_i$ to the loop $L_i$, our goal is to learn an approximation of the function $f(L, T) = S$. To this end, we train a neural network $f_p$ to minimize the mean squared error as our loss function:
$$\mathcal{L} = \frac{1}{N} \sum_{i=1}^{N} (f_p(L_i, T_i) - S_i)^2$$

\subsection{Preprocessing}

The input given to \Name{} is a set of loops, each extracted into a separate file from a larger program.
As discussed in Section~\ref{sec:background}, our dataset is based on loops used in the study by Gong et al.
Their technique for extracting loops can be easily applied to other programs as well.
Before training the model, we preprocess the data as follows. For each loop in the original program, we extract tokens from the source code, such that a token is represented as a pair $(t, v)$, where $t$ is its syntactic type and $v$ is the value, i.e., a string representation of the token in the source code.

For many learning tasks where the input data is a sequence of variable length it is common to select the maximum length beforehand. The input sequences of smaller lengths are then padded to the maximum length which makes it possible to vectorize the computations. To avoid long training times and be able to initialize the building blocks of our neural network, we exclude sequences of tokens longer than 250. In this way, we are able to achieve good model efficiency (Section~\ref{sub:efficiency_of_looplearner}) while keeping 90\% of the loops from the original dataset.


\subsection{Encoding of Source Code}
\label{sub:codeEncoding}

To feed the data to the neural network, we have to encode both the sequences of tokens and the transformations. The quality of encoding strongly impacts both the achievable level of accuracy and the generalization capability of the trained model.
We have experimented with multiple methods of encoding the sequences of tokens. Here we describe an interesting subset of these methods and their differences.

Some encoding methods are based on the frequency of tokens in the code corpus used for training.
Specifically, we compute the following three frequency maps:
\begin{itemize}
\item $F_{tokens}: \mathit{Token} \rightarrow \mathbb{N}$, which assigns a frequency to each token in the code corpus,
\item $F_{ids}: \mathit{Identifier} \rightarrow \mathbb{N}$, which assigns a frequency to each identifier in the code corpus,
\item $F_{stdTokens}: \mathit{Identifier} \rightarrow \mathbb{N}$, which assigns a frequency to each token that is neither an identifier nor a literal.
\end{itemize}

\begin{table*}
	\caption{Encoding methods for tokens.}
	\label{tab:codeEncoding}
  \setlength\tabcolsep{6pt} 
	\begin{tabular}{@{}p{4.8em}|p{3.5em}p{16em}p{13em}@{}}
		\toprule
		Encoding & Standard tokens & Identifiers & Literals \\
		\midrule
		Fixed & \multicolumn{3}{c}{One-hot encode top-n tokens, rest as \emph{unknown}} \\

		Basic & One-hot & All as \emph{id} & Keep integers, rest as \emph{unknown} \\

		Type-based & One-hot & Type of the identifier & Keep integers, rest as \emph{unknown} \\

		Renaming & One-hot & Consistent mapping to one-hot vectors & Keep integers, rest as \emph{unknown} \\

		Complex & One-hot & One-hot encoding of top c\%, rest as \emph{id} & One-hot encoding log(n) of integers, rest as \emph{unknown} \\

		FastText & \multicolumn{3}{c}{Learned embeddings of size 100} \\

		\bottomrule
	\end{tabular}
\end{table*}

Table~\ref{tab:codeEncoding} gives an overview of the six encoding methods we consider and which we explain in detail in the following.

\paragraph{Fixed encoding}
This encoding uses a one-hot encoding of the top $n$ most popular tokens in $F_{tokens}$ and assigns a special \textit{unknown} token to all other tokens.
This method is easy to implement, but has several disadvantages. First, the size of the encoding increases linearly with the size $n$ of the vocabulary, resulting in a increasing learning times. Next, all the words outside the vocabulary are encoded with the same unique token, which may result in a loss of vital information. Finally, this method does not discriminate between different types of tokens, i.e., keywords, identifiers, literals, etc.\ are all encoded as equidistant points in space.

\paragraph{Basic encoding}
This encoding is based on a one-hot encoding of all tokens in $F_{stdTokens}$, i.e., the set of standard tokens defined by the language, but not identifiers and literals. For literals, the encoding converts integer literals to base 10 and assigns special \emph{id} and \textit{unknown} tokens to identifiers and other tokens, respectively. In contrast to the fixed encoding, this method encodes the tokens based on their type.
The reason for handling integers specially is that we observe integers to sometimes influence optimization decisions, e.g., in loop headers.
In contrast, other literals, e.g., characters and floating-point values, are assumed not to influence the prediction accuracy and are therefore encoded as a special \emph{unknown} token. Omitting these tokens completely would change the structure of the code and potentially inhibit the performance of the neural network. The main disadvantage of this method is that it uses the same vector representation for all identifiers and thus hinders the learning capability of the network.

\paragraph{Type-based encoding}
This encoding is similar to basic, except that it replaces identifiers with the types of the corresponding variables for the most common data types: int, double, long, float, struct, char, short. While this method preserves the data type of many variables, all identifiers sharing the same data type get identical vector representations, which prevents the network from distinguishing them.

\paragraph{Renaming encoding}
This encoding is also similar to basic, except that each unique identifier is encoded as a one-hot vector of size $m$, where $m$ defines the maximum number of distinct identifier representations possible. The mapping from variable name to one-hot vector can be seen as a consistent renaming. This mapping is determined randomly, so as to prevent the order of the appearance of the identifiers from affecting the encoding.

Since the majority of unique tokens are identifier names, and because it is impractical to encode all identifiers, we use $D_i$ to calculate the minimum number of identifiers we would need to encode to cover a given percentage of tokens in the source code and store it in dictionary $I_{cov}$, where every integer percentage $p$ maps to the number of identifiers we would need to encode. Based on these statistics we devise various methods of encoding the data:

\paragraph{Complex encoding}
This encoding uses $F_{ids}$ to compute a minimal set of identifiers that covers at least c\% of all occurrences of identifiers across the code corpus.
Based on this set of frequent identifiers, the encoding preserves all frequent identifiers and only abstracts the remaining ones as \emph{unknown}.
Each integer literal is converted to a one-hot vector of size 64, based on the logarithm of its value. This is done to pass the scale of the literal to the network. In contrast to the fixed encoding and similar to the basic encoding, this method distinguishes among different token types, but also manages to cover a high number of unique identifiers.

\medskip
\noindent
The first five methods above encode tokens as one-hot vectors based on pre-calculated statistics. However, they all share the same disadvantages: the size of the vocabulary might become very large for big code corpora, and the tokens outside of the vocabulary are all represented as a single special \textit{unknown} token.
The following encoding addresses these limitations.

\paragraph{FastText encoding}
In natural language processing, an embedding~\cite{al2013polyglot, li2015multi, joulin2016bag, joulin2016fasttext} is a mapping of words to a vector of real numbers with a much lower dimension. It is a popular language modeling and feature learning technique already used for learning effective source-code representations~\cite{NIPS2018,oopsla2018-DeepBugs}. In our approach, we apply the FastText embedding technique~\cite{joulin2016bag} to source code.
We build FastText embeddings using all the sequences of token values in our training data. The size of the embedding vector is set to 100 and the model is trained for 100 epochs. Once this pre-training step is complete, we train our model by encoding token sequences with the help of the learned vector mappings for token values.
FastText is especially suitable for source code because many variable names are combinations of multiple words, for example, array\_size, viewCount, etc. Fasttext handles such names by not only learning embeddings for the tokens in the vocabulary but by also calculating embeddings for previously unseen words. This is done by breaking words into smaller sequences, calculating vector representations of each and using them to reconstruct the encoding of the whole word.

\subsection{Encoding of Code Transformations}
\label{sub:transformationEncoding}

To enable our model to learn effective transformations, we need to encode nearly 3,000 unique transformations with varying numbers of training samples for each transformation. A na\"ive approach is to use a one-hot encoding for all transformations. However, in this case, the size of the encoding vector would be very large and less popular transformations would not have enough associated data points for the training process to be successful. Furthermore, a one-hot encoding does not capture similarities between transformations, that is, all  transformations are represented as equidistant points in space, although some are much more similar than others. Another approach is to select only the most popular transformations and to one-hot encode them. While this allows us to train the model on the most common transformations, it has certain disadvantages. For example, by picking the 50 most popular transformations and ignoring the rest, we would lose 73\% of our data and therefore prevent our model from learning many beneficial transformations.

\begin{figure}[t]
\begin{center}
   \includegraphics[width=1\linewidth]{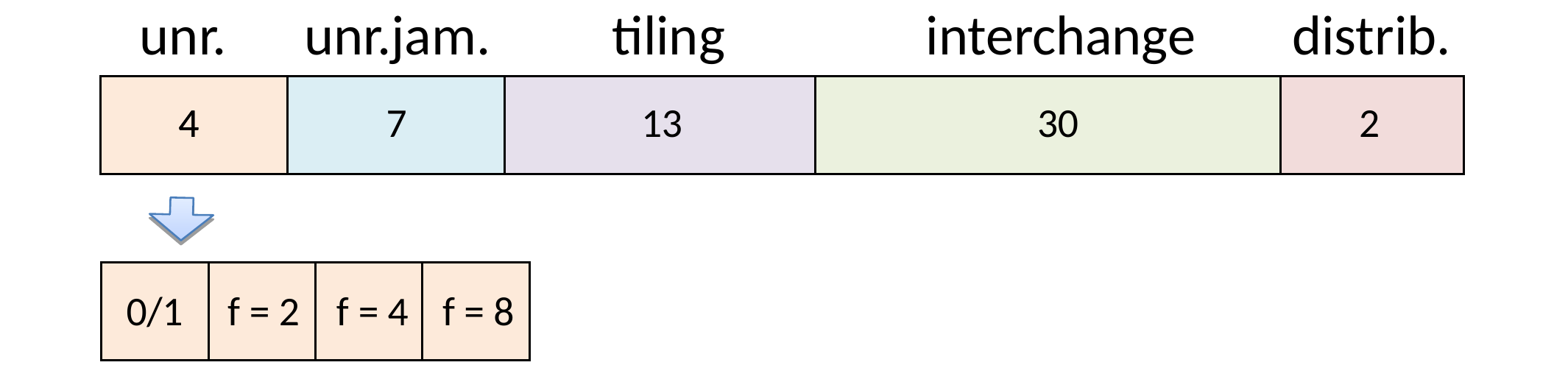}

\end{center}
   \caption{Vector encoding of transformations.}

\label{fig:t_encoding}
\end{figure}

To address the aforementioned points, we present \emph{compact} encodings of code transformations, where each sequence of transformations is represented as a feature vector.
The encoding exploits the fact that transformations can only be applied in particular orders that preserve the semantics of the original program (Section~\ref{sec:background}).
Because the set of transformations included in a sequence of transformations is sufficient to uniquely specify the sequence, the features in the encoding indicate the presence or absence of a particular transformation and the set of its parameters.
We formally define the encoding as follows.
\begin{definition}[Compact encoding of transformations]
We encode a sequence of transformations $T$ as a concatenation of vectors $T_1$, .. , $T_k$, where each $T_i$ represents a vector encoding for transformation $i$. The size of a vector $T_i$ is equal to a maximum number of different parameterizations of transformation i. The first element in $T_i$ indicates whether $i$ is applied, while the subsequent elements indicate which parameter of $i$ is enabled.
For the transformations considered in this work, we define the size of subvectors $T_i$ as follows:
\begin{itemize}
\item $size(T_{unroll}) = 4$
\item $size(T_{unrolljam}) = 7$
\item $size(T_{tiling}) = 13$
\item $size(T_{interchange}) = 30$
\item $size(T_{distribution}) = 2$
\end{itemize}
\end{definition}

Figure~\ref{fig:t_encoding} illustrates the compact encoding of loop transformations. The final size of the encoding vector is 56. The first four elements are reserved for the $T_{unroll}$ subvector. The first element in $T_{unroll}$ has a value of 0 or 1, indicating whether  unrolling is part of the transformation (0-no, 1-yes). The next three elements are used to encode the unrolling factor. For example, if unrolling is applied with factor 2, then the first two elements of $T_{unroll}$ would have value 1 and the remaining ones would be set to 0. We encode other transformations in a similar fashion, taking into account all possible combinations of their parameters.

\subsection{DNN Architecture}
\begin{figure}
\begin{center}
  \includegraphics[width=1\linewidth]{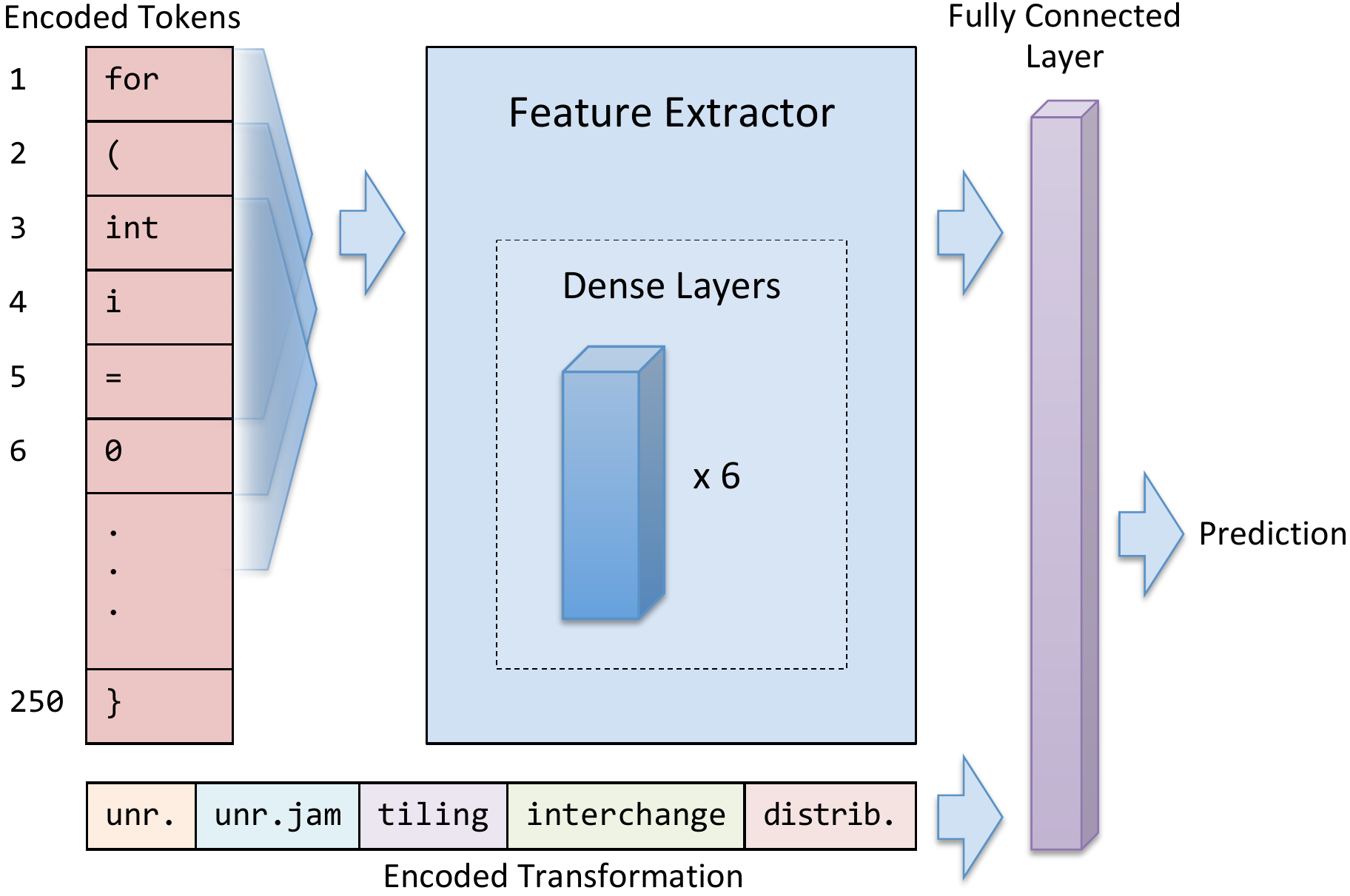}
\end{center}
   \caption{Prediction process for a sequence of tokens and transformations. The encoded sequence of tokens is first passed into the feature extractor. The results are concatenated with the encoded vector of transformations and passed to the fully connected layer which predicts the speedup.}

\label{fig:dnn_architecture}
\end{figure}

\begin{figure}
\begin{center}
  \includegraphics[width=.75\linewidth]{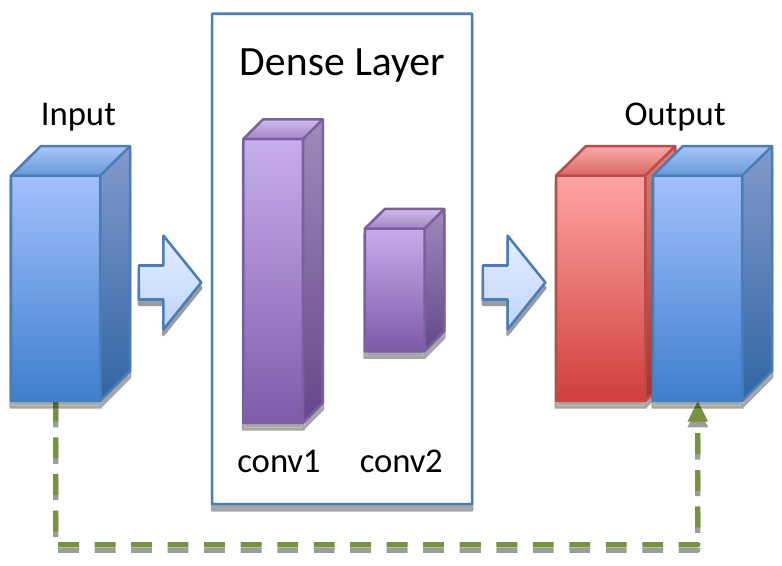}
\end{center}
   \caption{The dense layer of the DNN consists of two convolutional layers. The outputs of the dense layer are concatenated with the inputs and passed on to the next layer.}

\label{fig:dense_layer}
\end{figure}
To train a model that predicts beneficial transformations for a loop, we consider two different network architectures: \emph{recurrent} and \emph{convolutional}. Recurrent neural networks (RNN) have been designed to recognize patterns in sequences of data, such as text or numerical times series. The main property of RNNs is the internal memory used to keep outputs of the previous steps, which is then fed as input to the current step. In contrast, convolutional neural networks (CNN) are suitable for hierarchical data. The most distinctive property of CNNs are their \emph{convolutional layers}, which perform a mathematical \emph{convolution} operation on the input data. Convolutional layers consist of feature matrices that learn to recognize features in the input. Stacking convolutional layers on top of each other allows the later layers to learn increasingly complex features, which makes CNNs so powerful for any task involving compositional data.

The advantage of recurrent neural networks is that they process sequences of arbitrary length. However, vanishing gradients and increased computational demand for the training process makes it harder to train the network with very long input sequences. While convolutional neural networks lack the ability to process sequences of variable length, they excel on datasets of compositional data. Since the source code is not only sequential but also highly compositional, CNNs are a good fit for this task.
Our experimental evaluation shows that convolutional networks have a higher level of accuracy compared to recurrent networks. This is why we decided to choose CNNs as our default architecture.

Specifically, we adopted ideas from DenseNet~\cite{Huang2016}, a well-known design from the field of computer vision. However, we custom-tailored the DenseNet architecture to fit our learning problem. Figure~\ref{fig:dnn_architecture} further illustrates the architecture of our model. The inputs to our model are encoded tokens of a loop and encoded transformations. Because our input is compositional along a single dimension, we use one-dimensional convolutions instead of the two-dimensional variants used in the original DenseNet. The building blocks of our neural network are dense layers which learn to extract features from the source code. As further illustrated in Figure~\ref{fig:dense_layer}, each dense layer consists of two convolutional layers and the inputs of each dense layer are concatenated with the outputs of previous layers and fed into subsequent layers. Eventually, the model performs average-pooling on the outputs of the final convolutional layer, concatenates the results with the transformation vector, and passes the concatenated vector into a fully connected layer, which is used to predict the expected speedup.

\subsection{Training}

We feed training samples in batches of 256 into the network and use the stochastic gradient descent method to train the network for 300 epochs. The initial learning rate of 0.001 is dropped to a third of its value in epochs 100 and 200, and a momentum factor of 0.9 is used for optimization. We clip the gradients with the absolute value above 10 to avoid the exploding gradients problem. At the end of every epoch, we evaluate the model and save the best-performing model.

\subsection{Implementation}
\label{sec:implementation}

The code is parsed and tokenized by using the lexer component of the Python \emph{pycparser}\footnote{\url{https://github.com/eliben/pycparser}} library, a parser for the C language.
To build and train the models we use the \emph{PyTorch} framework, version 0.4.1\footnote{\url{https://pytorch.org/docs/0.4.1/}}. We implement \Name{} as an extensible framework that takes as input the following key parameters:
\begin{itemize}
  \item \emph{sequence encoding}: fixed, basic, type-based, renaming, complex, or fasttext
  \item \emph{transformation encoding}: one-hot or compact
  \item \emph{model type}: recurrent or convolutional
\end{itemize}

This allows easy plugin of new types of encodings and neural network architectures.

\section{Evaluation}
\label{sec:eval}

%
%
%
%
%
%
%


Our evaluation focuses on the following questions.
\begin{itemize}
	\item How effective is \Name{} at predicting beneficial loop transformations? (Sections~\ref{sub:overall_accuracy}, \ref{sub:topk_accuracy}, and~\ref{sub:successful_transformation_combinations})
	\item What speedups do \Name{}'s predictions enable? (Section~\ref{sub:speedups})
	\item How efficient is \Name{}? (Section~\ref{sub:efficiency_of_looplearner})
	\item How does the approach compare to exhaustively trying all loop transformations? (Section~\ref{sub:comparison_with_exhaustive_search})
	\item What is the influence of the speedup threshold? (Section~\ref{sub:tuning_speedup_threshold})
\end{itemize}

\subsection{Experimental Setup}

Our dataset is built from 1,895 base loops extracted by prior work~\cite{Gong:2018} from various benchmarks, software
libraries, and machine-learning kernels written in C.
Extracting each loop into a standalone program that replicates the data environment of the original benchmark program, applying sequences of transformations, and measuring their performance yields a dataset of roughly 70,000 (loop, transformation, speedup) triples.
The loops are compiled with the GNU GCC compiler, using the -O3 flag, and
executed on an Intel Xeon E5-1630 v3 processor.

We split the dataset into a training and a validation set by randomly selecting  80\% of all loops and their associated transformations for training, and the remainder for validation.
By splitting by loop, we ensure that the evaluation measures how well the approach performs on previously unseen loops. Unless explicitly stated otherwise, we use speedup threshold $t=1.0$.
We trained our models on a single server with two Intel(R) Xeon(R) Gold 6126 2.60GHz CPUs, 64GBs of main memory, two NVIDIA GeForce GTX 1080 Ti GPUs, and Ubuntu 16.04 LTS operating system. For the purpose of training any given model, a single GPU was used at a time.

\subsection{Overall Accuracy of Predictions}
\label{sub:overall_accuracy}

\subsubsection{Metrics}
We first measure the accuracy of \Name{}'s predictions across all loops and transformations in the validation set.
Let $T$ be the set of all (loop, transformation) pairs. Let $T^{+} \subseteq T$ and $T^{-} \subseteq T$ be the subset of all pairs known to cause a speedup and slowdown, respectively. Let $P^{+} \subseteq T$ and $P^{-} \subseteq T$ be the subset of all the pairs predicted to result in a speedup and slowdown, respectively. We consider the following metrics:

\begin{itemize}
	\item \emph{Total accuracy (\%)} is the percentage of elements out of $T$ that are in ($P^+ \cap T^+) \cup (P^- \cap T^-$).
	\item \emph{Speedup recall (\%)} is the percentage of elements out of $T^+$ that are in $P^+ \cap T^+$.
	\item \emph{Speedup precision (\%)} is the percentage of elements out of $P^+$ that are in $P^+ \cap T^+$.
	\item \emph{Slowdown recall (\%)} is the percentage of elements out of $T^-$ that are in $P^- \cap T^-$.
	\item \emph{Slowdown precision (\%)} is the percentage of elements out of $P^-$ that are in $P^- \cap T^-$.
\end{itemize}

We calculate the last four metrics alongside the total accuracy for two reasons.
First, our dataset is imbalanced---more than 80\% of transformations result in slowdown and therefore high total prediction accuracy alone does not necessarily imply high accuracy for both speedups and slowdowns.
Second, the recall and precision metrics help understand how well the approach performs in a particular usage scenario.
For example, speedup precision shows how often a predicted speedup indeed improves the loop's performance.
We also show the F1 score (harmonic mean of precision and recall).

\subsubsection{Results}
\label{subsub:overall_results}
Table~\ref{tab:overall_experiments} summarizes the results.
To understand the influence of different encodings and models, we report results for different variants of \Name{}.
The best result for each metric is highlighted in bold font.
Overall, the approach predicts beneficial loop transformations with high accuracy (up to 88\%).
Comparing speedup and slowdown predictions, the model is particularly effective at predicting that a transformation will cause a slowdown (95\% recall, 92\% precision), but also provides reasonable results for speedups (55\% recall, 66\% precision).

\begin{table*}
	\centering
	\caption{Overall accuracies achieved by employing different encoding methods. Training accuracy reflects the highest achieved accuracy on the training set. All other values refer to the validation set.}
  \setlength\tabcolsep{10pt} 
	\begin{tabular}{@{} l | c c | r r r | r r r @{}}
		\hline
		& \multicolumn{2}{ c |}{Accuracy (\%)} & \multicolumn{3}{ c |}{Speedup (\%)} & \multicolumn{3}{ c }{Slowdown (\%)} \\
		\hline
		Sequence Encoding & \makecell{Training} & \makecell{Validation} & \makecell{ Recall } & \makecell{Precision} & F1 & \makecell{Recall} & \makecell{Precision} & F1 \\
		\hline
		\multicolumn{3}{ l }{\textit{Transformation Encoding: Compact}} & \multicolumn{4}{ l }{\textit{Model: CNN}} \\
		\hline
		Fixed(n=1,000)  &  \textbf{92.5} & 87.6 & 55.9  & 63.0 & 59.2 & 93.7 & 91.7 & 92.7 \\
		Basic          &  90.0          & 84.0 & 7.7   & 54.8 & 13.5 & 98.8 & 84.7 & 91.2 \\
		Type-based     &  89.8          & 84.0 & 4.3   & 56.4 & 7.9 & 99.4 & 84.3 & 91.2 \\
		Renaming(m=40)   &  88.8          & 84.0 & 11.1  & 51.7 & 18.3 & 98.0 & 85.1 & 91.1 \\
		Complex(c=70\%)&  92.1          & 87.9 & 57.0  & 64.1 & 60.4 &93.8 & 91.9 & 92.9 \\
		Complex(c=80\%)&  92.0          & 87.7 & 58.2  & 62.9 &\textbf{60.5} & 93.4 & 92.1 & 92.7 \\
		FastText       &  92.0          & \textbf{88.1} & 54.8 & 66.1 & 59.9 & 94.6 & 91.6 &\textbf{93.0} \\
		\hline
		\multicolumn{3}{ l }{\textit{Transformation Encoding: One-hot}} & \multicolumn{4}{ l }{\textit{Model: CNN}} \\
		\hline
		FastText & 89.2 & 87.1 & 43.0 & 65.3 & 51.8 & 95.6 & 89.7 & 92.5 \\
		\hline
		\multicolumn{3}{ l }{\textit{Transformation Encoding: Compact}} & \multicolumn{4}{ l }{\textit{Model: RNN}} \\
		\hline
		FastText & 84.8 & 84.0 & 4.4 & 56.0 & 8.1 & 99.3 & 84.3 & 91.2 \\
		\hline
	\end{tabular}
	\label{tab:overall_experiments}
\end{table*}


\paragraph{Comparison of source code encodings}

Remarkably, fixed encoding achieves the highest accuracy on the training set and relatively good accuracy on the validation set, while also being the easiest to implement.
We attribute this result to the higher dimensionality of the input data. Since each token is represented as a vector in space $\mathbb{R}^{1001}$, that is, each of the top 1,000 most common tokens and a special \textit{unknown} token get unique representations, it is quite easy for the network to learn to differentiate between distinct tokens. However, apart from the size of the input data, the disadvantage of using fixed encoding when compared to more advanced methods is that the gap between the training and validation set accuracy for this method is also quite high, which means it tends to overfit the training data while not performing as well on the validation set. The reason is that the top 1,000 most common tokens are extracted from the training set, which is likely to be somewhat different from the validation set.

Although the accuracy achieved by the ``basic'' encoding is roughly that of other encodings, the speedup prediction results show a significant weakness of the ``basic'' encoding. The model achieves only 7.7\% speedup recall, because crucial information is lost when discarding identifier names, float literals, and char literals during encoding. As shown by the results of the ``type-based'' encoding, replacing identifier names with type information does not make the model any more accurate. Consistently abstracting variable names into generic names (``renaming'') slightly improves the results, but still offers only low speedup prediction results.
The main take-away of these results is that identifier names and literal values are helpful in learning-based program analysis, a finding in line with other work on name-based and learning-based analysis~\cite{oopsla2018-DeepBugs,icse2019}.

The ``complex'' encoding method achieves fairly high training- and validation-set accuracy.
The substantially higher accuracy compared to ``basic'' encoding confirms the importance of encoding identifier names. However, comparing the two variants of ``complex'', which keep 70\% and 80\% of all identifiers, respectively, shows that adding another 10\% of less common identifier names does not raise the accuracy any further. We believe that after a certain point, increasing the size of the encoding vector by adding rare identifier names does not benefit the accuracy of the trained model and can actually be harmful, since it is likely that the model will learn to overfit the training samples based on the occurrence of rare identifiers.

The ``FastText'' encoding achieves the highest overall validation accuracy, showing that pre-training general-purpose token embeddings before passing them into a task-specific model is beneficial. The difference between training accuracy and validation accuracy is at a minimum when using the ``FastText'' encoding, i.e., there is only little overfitting.
Since we obtain the best overall accuracy the ``FastText'' encoding, this encoding is the default in the remainder of the section.

\paragraph{Comparison of transformation encodings}
Comparing our compact encoding of transformations with a naive one-hot encoding of transformations shows that the compact encoding is beneficial.
In particular, it enables the model to predict otherwise missed speedups.
We attribute this result to the fact that the dense encoding makes it easier for the model to generalize across similar transformations, as those are encoded into similar vectors.

\paragraph{Comparison of neural architectures}
We compare our default CNN-based neural architecture to a recurrent neural network with two layers of gated recurrent units and a size similar to the CNN architecture.
The comparison shows the CNN model to be clearly more effective, in particular in predicting speedups.

\subsection{Effectiveness of Top-k Predictions per Loop}
\label{sub:topk_accuracy}

\subsubsection{Metrics}

To better understand how effective \Name{} is for individual loops, we evaluate the effectiveness of those $k$ transformations per loop that \Name{} predicts to have the highest speedups.
Let $L$ be the set of all the loops, let $L^+ \subseteq L$ be the subset of the loops for which there exists at least one transformation that produces a speedup, let $P_o^{(l)}$ be the set of transformations that can be applied to the loop $l \in L$ ordered by the predicted performance from highest to lowest, let $P_o^{(l)}(k) \subseteq P_o^{(l)}$ be the first $k$ transformations in this set, let $P_{os}^{(l)}(k) \subseteq P_o^{(l)}(k)$ be the subset of transformations that are predicted to be advantageous, and let $L_{sp} \subseteq L$ be the subset of the loops for which $P_{os}^{(l)}(1) \ne \varnothing$. Then, to measure the top-k effectiveness of our model we calculate:
\begin{itemize}

	\item \emph{Total accuracy (\%)} is the percentage of loops $l \in L$ for which at least one of the predictions for transformations in $P_o^{(l)}(k)$ is correct.

	\item \emph{Speedup recall (\%)} is the percentage of loops $l \in L^+$ for which at least one transformation in $P_{os}^{(l)}(k)$ produces a speedup.

	\item \emph{Speedup precision (\%)} is the percentage of loops $l \in L_{sp}$ for which at least one transformation in $P_{os}^{(l)}(k)$ produces a speedup.
\end{itemize}

\subsubsection{Results}

Table~\ref{tab:top_k_experiments} shows the results (the last two columns are described later).
We find that the approach achieves an accuracy of 65\% when considering only the top-most prediction for a loop, and of 83\% within the top-5 predictions.
The precision of speedups ranges between 73\% and 76\% percent, i.e., when the model predicts a speedup, then the code indeed performs faster in most cases.
The reason why the validation accuracy for top-1 predictions is lower than the overall accuracy is that the distribution of the numbers of possible transformations across the loops is non-uniform.
Some loops have a much higher number of valid transformations than others, and for some loops the top-1 prediction is more likely to be accurate than for others.

\begin{table}
\centering
\caption{Top-1, top-3 and top-5 accuracy of the network on the validation set and the corresponding values for precision, recall, and the mean speedup achieved in both static and dynamic mode of execution.}

\setlength\tabcolsep{3pt}
\begin{tabular}{ c | c | c c | c c }
 \hline
  Top & Total & \multicolumn{4}{ c }{Speedup} \\
 \hline
 k & Acc. (\%) & Recall (\%) & Precision (\%) & Static & Dynamic \\
 \hline
 1 & 64.91 & 39.46 & 73.05 & 1.144x & 1.235x \\
 3 & 79.95 & 40.61 & 75.18 & N/A    & 1.285x \\
 5 & 83.38 & 41.00 & 75.89 & N/A    & 1.290x \\
 \hline
\end{tabular}
\label{tab:top_k_experiments}
\end{table}

\subsection{Speedups Achieved due to \Name{}}
\label{sub:speedups}

\subsubsection{Metrics}
\label{sub:speedupsMetrics}
We evaluate the speedups obtained by applying the transformations suggested by \Name{} in both the static and the dynamic usage scenario (Section~\ref{sub:usageScenarios}).
The speedups in the static scenario show the performance improvement that can be immediately achieved when applying \Name{}'s top suggested transformations, while the dynamic scenario shows the potential speedup attainable when validating \Name{}'s predictions.
We compute the following two metrics:
\begin{itemize}
\item \emph{Speedup geometric mean (static)} is defined only for $k = 1$ and is the geometric mean of speedups across all loops $l \in L_{sp}$ achieved when applying transformation $P_{os}^{(l)}(1)$.

\item \emph{Speedup geometric mean (dynamic)} is the geometric mean of speedups across all loops $l \in L_{sp}$ achieved when applying the transformation with the best performance out of $P_{os}^{(l)}(k)$, or 1.0 if none of the top-$k$ transformations results in speedup.
\end{itemize}

\subsubsection{Results}
The last two columns of Table~\ref{tab:top_k_experiments} shows the speedups for both scenarios.
We find that \Name{} enables significant speedups in both cases, with a 1.14x speedup when simply using the top-1 prediction, and an 1.29x speedup when choosing the best from the top-5 predictions.
Because in the dynamic scenario, the transformed loops are executed to measure their performance, the mean speedup is guaranteed to be at least as high as in the static scenario.

\subsection{Efficiency of \Name{}}
\label{sub:efficiency_of_looplearner}

We summarize the execution time for different phases of our approach when running on either CPU or GPU in Table~\ref{tab:performance_breakdown}. Before training our model we learn FastText embeddings, which takes about 20 seconds on our dataset using 32 worker threads. By far the most computationally demanding part of our approach is training the neural network. With hyperparameter settings discussed earlier it takes around 6 hours and 40 minutes to complete the training. However, we believe this time can be brought down substantially by using higher batch sizes along with more memory-efficient implementations of the DenseNet architecture. Moreover, the training step, despite being the most time-consuming, is only performed once and afterwards the resulting model is ready to be deployed.

Because a high number of transformations can be applied to a given loop, our model must be executed many times before it is possible to decide which transformation is the most beneficial. During prediction, the most computationally intensive part is the feature extractor, which processes the token sequences of a loop. Fortunately, it is sufficient to run the feature extractor for any given loop only once. Then, the fully connected layer can be used to evaluate many possible transformations in a batch. As can be observed in Table~\ref{tab:performance_breakdown}, it takes less than 20 milliseconds to evaluate 1,000 transformations for a single loop on a CPU and less than 2 milliseconds for the same task on a GPU. We believe that these results show that it is practical to implement \Name{} as an automated pre-processing step before giving code to the compiler.

\begin{table}
	\centering
	\caption{Time requirements of the different phases of our approach.}
	\setlength\tabcolsep{1.5pt}
	\begin{tabular}{ l | l | l }
		\hline
		Approach phase & Time (CPU) & Time (GPU) \\
		\hline
		Learning embeddings & 20 seconds & N/A \\
		Training (1 epoch) & N/A & 60 seconds \\
		Evaluation (single pass) & N/A & 20 seconds \\
		Full training (300 epochs) & N/A & 6.6 hours \\
		Evaluating 1 transformation & 13.0 ms & 1.6 ms \\
		Evaluating 100 transformations & 13.5 ms & 1.6 ms \\
		Evaluating 1,000 transformations & 15.9 ms & 1.7 ms \\
		\hline
	\end{tabular}
	\label{tab:performance_breakdown}
\end{table}

\subsection{Comparison with Exhaustive Search}
\label{sub:comparison_with_exhaustive_search}

An alternative to querying \Name{} for transformations that are likely to improve the performance of a loop is exhaustive search through all possible sequences of transformations.
By measuring the performance impact of each sequence of transformations, that alternative approach is guaranteed to always find the best-performing representation of a loop.
The downside is that it is very time-consuming, as repeatedly executing different variants of a loop takes time.
The following explores the trade-off between time spent on finding beneficial transformations and time saved during the loop executions.

\paragraph{Time to find beneficial transformations}
It takes about 10 hours to exhaustively measure the runtime of all mutations in our dataset.
This time is based on executions of individual loops extracted from their original program~\cite{Gong:2018}, and it excludes the time required for extracting the loops.
In contrast, predicting the speedup of transformations across all loops using our model takes less than 2 seconds.
\Name{} hence reduces the time taken to select a suitable transformation by multiple orders of magnitude.

\paragraph{Time savings due to optimized loops}
We compare \Name{} and exhaustive search w.r.t.\ the speedup obtained across all loops for which the respective approach suggests applying a transformation.
For \Name{}, those are all loops for which at least one transformation is predicted to yield a speedup.
For exhaustive search, those are all loops that actually have at least one such transformation.
Intuitively, the speedup hence indicates what benefits to expect when following the suggestions of the two approaches.
As shown in Table~\ref{tab:top_k_experiments}, \Name{}'s static usage scenario yields a speedup of 1.144x.
In contrast, exhaustive search yields a speedup of 1.286x.
That is, following the top-most suggestion of the model without validating its performance impact results in lower but still relevant speedups.
\Name{}'s dynamic usage scenario shows a different picture.
By considering the top-5 suggestions of the model, the obtained speedup of 1.290x even exceeds that of exhaustive search.
The reason is that exhaustive search also reveals various transformations that yield very small speedups, i.e., transformations that are less relevant in practice.
Intuitively, the top-5, dynamic scenario can be seen as an exhaustive search within a much reduced space of only the five most promising transformations.

\medskip \noindent
Overall, we conclude that \Name{} provides a practical alternative to exhaustive search, allowing developers or automated tools to quickly identify the most beneficial loop optimizations.
In particular, the dynamic mode identifies many of those transformations that yield a significant speedup, without paying the cost of exhaustively measuring the performance impact  of all transformations for all loops.

\subsection{Successful Combinations of Transformations}
\label{sub:successful_transformation_combinations}

To better understand for which transformations the model's predictions are more or less accurate, Table~\ref{tab:transformation_combinations} shows results for individual sequences of
transformations.
The abbreviations for the transformations are as in Table~\ref{tb:parameters}.
The last two columns show the number of loops in the validation set to which a sequence of transformation applies, and what percentage of the validation set this number comprises (i.e., coverage).
The results show that the accuracy varies across transformations.
For example, tiling followed by unrolling has a relatively low validation accuracy, but a high training accuracy, which indicates that the model has likely overfit the training data for this transformation sequence.
We also observe that for some under-represented combinations of transformations, the model fails to identify a single speedup.
By observing the results for individual transformation sequences, one might decide to ignore some sequences when deploying \Name{}.

\begin{table*}
\centering
\caption{Performance of the neural network on different sequences of transformations. Precision and recall for speedup are calculated on the validation set.}
\setlength\tabcolsep{8pt}
\begin{tabular}{ l | c c | r r | r r }
 \hline
   & \multicolumn{2}{ c | }{Accuracy (\%)} & \multicolumn{2}{ c |}{Speedup (\%)} & \multicolumn{2}{ c }{Loop Coverage} \\
 \hline
  Transformation Sequence & Training & Validation & Recall & Precision & Count & \% \\
 \hline
 unrolling                                 & 66.75& 60.30& 35.49& 70.33& 379&100.00\\
 tiling                                  & 80.44& 71.08& 14.20& 53.33& 156& 41.16\\
 tiling -> unrolling                       & 82.76& 66.14& 10.68& 45.67& 156& 41.16\\
 unroll-and-jam -> unrolling                    & 71.00& 69.25& 27.33& 87.23&  46& 12.14\\
 interchange                                  & 90.15& 89.40& 50.98& 78.79&  46& 12.14\\
 interchange -> unrolling                       & 93.69& 89.37& 43.88& 88.41&  46& 12.14\\
 interchange -> unroll-and-jam                     & 92.37& 91.02& 53.28& 77.71&  46& 12.14\\
 interchange -> unroll-and-jam -> unrolling          & 92.91& 92.81& 54.15& 83.19&  46& 12.14\\
 interchange -> tiling -> unrolling             & 94.52& 93.15& 20.65& 72.52&  44& 11.61\\
 interchange -> tiling                        & 93.28& 93.26& 22.22& 67.92&  44& 11.61\\
 distribution                                & 69.47& 54.55& 63.64& 53.85&  22&  5.80\\
 distribution -> unrolling                     & 74.64& 55.56& 41.18& 63.64&  22&  5.80\\
 tiling -> distribution -> unrolling           & 85.45& 78.53& 17.07& 63.64&  16&  4.22\\
 tiling -> distribution                      & 81.62& 69.49&  7.14& 16.67&  16&  4.22\\
 interchange -> distribution                      & 96.55& 77.78&  0.00&  0.00&   5&  1.32\\
 interchange -> distribution -> unrolling	         & 98.13& 84.62& 20.00&100.00&   5&  1.32\\
 interchange -> tiling -> distribution            & 97.51& 94.92&  0.00&  0.00&   4&  1.06\\
 interchange -> tiling -> distribution -> unrolling & 98.82& 94.92&  0.00&  0.00&   4&  1.06\\
 \hline
\end{tabular}
\label{tab:transformation_combinations}
\end{table*}

\subsection{Influence of Speedup Threshold}
\label{sub:tuning_speedup_threshold}

\begin{figure}[t]
\begin{center}
   \includegraphics[width=1\linewidth]{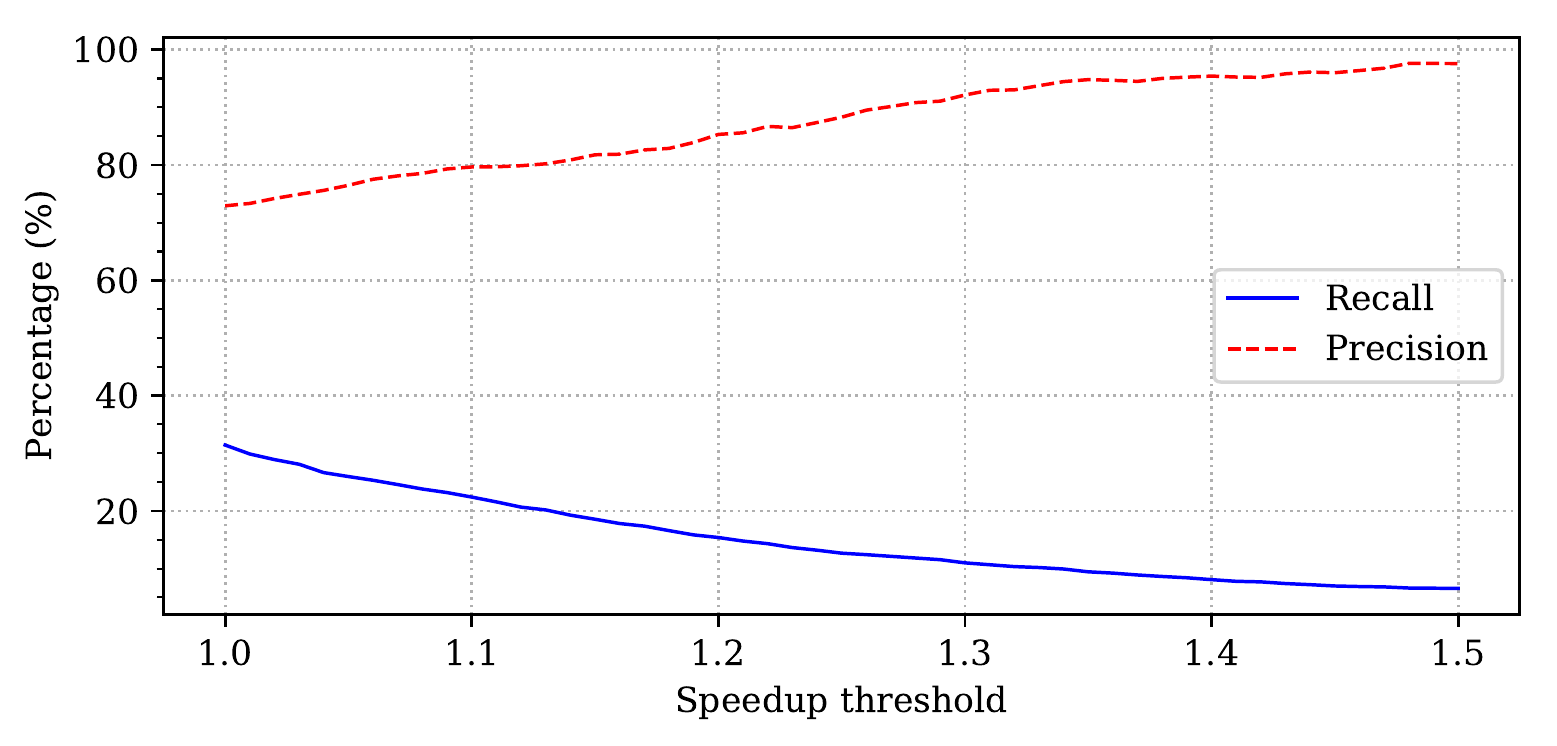}
\end{center}
   \caption{The effect of the speedup threshold on the validation-set speedup precision and recall.}

\label{fig:threshold_precision_recall}
\end{figure}

So far in our evaluation we defined the speedup threshold as being equal to 1. However, as mentioned earlier, this hyperparameter can be used to adjust the precision and recall of the trained model. To show the effects of tuning this hyperparameter, we evaluate the speedup precision and recall on the validation set as we increase the speedup threshold from 1.0 to 1.5. Figure~\ref{fig:threshold_precision_recall} shows that, predictably, increasing the speedup threshold will result in higher precision of speedup predictions but also reduce the recall percentage. Lower value settings for this hyperparameter might be suitable for a more optimistic approach with high tolerance for speedup mispredictions. On the other hand, higher values guarantee a lower number of mispredictions but are also likely to disregard advantageous transformations producing smaller speedups.

\section{Related Work}
\label{sec:related}



Since many compiler bugs are triggered by optimizations~\cite{Chen:2016}, several techniques search for optimization-related compiler bugs via differential testing~\cite{Yang:2011, Le:2014}.
\citet{Barany2018} compare the code
generated by different compilers to find optimizations performed by
one but missed by another compiler.
Similarly, \citet{Nagai2014, Nagai2019} propose testing the validity of arithmetic optimizations using randomly generated programs.
Instead of searching for bugs in the implementation of compiler optimizations,
our work improves the effectiveness of optimizations by tailoring
loops to the optimization decisions made by the compiler.


Superoptimization tries to find the best program among all semantics-preserving variants of a given program~\cite{massalin1987superoptimizer} and can, e.g., be addressed as a stochasitic search problem~\cite{Schkufza2013}.
\citet{Bunel2017} propose a learning-based approach to improve superoptimization by predicting the distribution of code transformations to sample from.
Another search-based approach for finding suitable optimizations is evolutionary search, e.g., to tune the order of optimizations~\cite{Cooper1999,Cooper2002},
to decide which optimizations to enable~\cite{Hoste2008},
or to apply random code mutations that reduce energy consumption~\cite{Schulte2014}.
All of the above approaches search the optimization space for a specific program and pay the cost, e.g., for executing and validating candidate programs, for every program.
In contrast, \Name{} learns a model once, which then predicts code transformations suitable
for the given program without the need to execute or validate candidate programs.
A difference to the work by \citet{Cooper1999}, which also looks for sequences of code transformations, is that their work optimizes in which order to apply transformations, whereas our work predicts whether applying any transformation will be beneficial, and if yes, which sequence of transformations to choose.


\citet{Monsifrot:2002} use decision trees to learn
the behavior of loop unrolling optimizations to decide which loop to
unroll. \citet{Stephenson:2005} propose a supervised learning
algorithm to predict unroll factors. \citet{Yuki:2010}
train a neural network to predict loop tiling
sizes. \citet{Simon:2013} automatically learn effective inlining
heuristics using decision trees and static code features.
Machine learning has been also applied to predict an effective
application order of compiler optimizations~\cite{fursin2011milepost, Park:2012,
  Martins:2016, Ashouri:2017}. \citet{Park:2012} use a
graph-based intermediate representation to train a model that predicts optimization sequences that will benefit a given
program. \citet{Martins:2016} propose a clustering approach for
grouping similar functions, reducing the search space resulting from
the combination of optimizations previously suggested for the
functions in each group. \citet{Ashouri:2017} cluster compiler
optimizations to predict the speedup of sequences of optimizations that belong to the same cluster.
All the above approaches differ from our work by tuning optimization
decisions made inside the compiler, whereas we present a
pre-processing step that makes optimizations more efficient without
changing the compiler itself.  Another difference is that the above
methods rely on manually designed features.

Recent work by Cummins et al.~\cite{Cummins2017, Cummins2017a} also proposes a deep neural
network that learns optimization heuristics over raw code, similar to
our work. Their work focuses on heuristics for two optimization
problems: predicting the optimal execution device and the thread
coarsening factor.  Our work differs in at least two ways. First,
\Name{} learns effective transformation sequences from a much larger
corpus of transformations. Second, \Name{} trains a convolutional
neural network, whereas Cummins et al.\ build upon a recurrent neural
network.
Another technique optimizes the memory layout of matrices to enable faster sparse matrix multiplication~\cite{Zhao2018c}.
While also being based on convolutional neural networks, their approach takes a matrix as the input, whereas \Name{} reasons about the code to optimize.


%


Machine learning has been used to address various programming-related
problems in an end-to-end manner~\cite{Allamanis2018}, including code
completion~\cite{Nguyen2013a,Raychev2014,Bielik2016}, bug
detection~\cite{oopsla2018-DeepBugs,Li2018a,Harer2018a}, and bug
fixing~\cite{Long2016,Gupta2017,Harer2018}.
%
%
Recurrent neural networks have been applied to token sequences, for
example, to find fixes for syntax errors~\cite{Bhatia2016}, to
identify code that suffers from a specific kind of
vulnerability~\cite{Li2018a}, to predict the types of
variables~\cite{Hellendoorn2018}, or to represent code for code
search~\cite{Gu2018}.  An alternative to recurrent neural networks are
convolutional networks, which we also use in this paper.  Others have
used convolutional networks to localize bugs~\cite{Huo2016} and
to summarize code~\cite{Allamanis2016}.
We address a different prediction problem, and we are, to the best of our
knowledge, the first to adopt the DenseNet
architecture~\cite{Huang2016} to code.
%
%
Several techniques train models using a graph-based code representation, e.g., abstract syntax trees~\cite{White2016,Zhang2019}, paths through abstract syntax trees~\cite{Alon2018,Alon2018a,Devlin2017}, control flow graphs~\cite{DeFreez2018}, execution
trees~\cite{Henkel2018}, and other graph-based code representations~\cite{NIPS2018,Allamanis2017b,Brockschmidt2018,Xu2017}.
Other models of code build on conditional random fields~\cite{Raychev2015}, memory networks~\cite{Choi2017}, or manually modeled features~\cite{Zhao2018}.
We build upon a token sequence-based representation instead, because it
is conceptually simple and makes training efficient, while providing
accurate predictions.
%
%

\section{Conclusion}
\label{sec:con}

We present \Name{}, a novel technique to address the program of compiler instability.
Given the source code of a loop, LoopLearner
suggests a semantically invariant transformation that will likely
allow the compiler to produce more efficient code. Following its
recommendations prior to compilation results in an average speedup of
1.14x. Almost three quarters (73\%) of the suggested transformations
yield positive speedups. Trying the top-3 recommendations and choosing
the best one raises the average speedup even to 1.29x.
We envision the approach to be used either as a tool to guide programmers or as a pre-processor run before or as part of the compiler. Different from most earlier work, our approach
leverages deep learning and does not require any manual selection of
source code features. In addition, we consider a much larger set of
transformations---3,000 combinations of five common loop optimizations
in our case. Our model needs to be trained once per compiler and
platform, an effort that is likely to pay off in view of the typical
lifetime of either of the two.

\section*{Acknowledgments}

This work was supported by the Graduate School CE within the Centre for Computational Engineering at Technische Universität Darmstadt, the Hessian LOEWE
initiative within the Software-Factory 4.0 project, European Research Council (ERC, grant agreement 851895), and the German Research Foundation within the
ConcSys and Perf4JS projects.

\bibliography{bibfile}

\appendix

\end{document}